\documentclass[conference]{IEEEtran}
\IEEEoverridecommandlockouts
\usepackage{cite}
\usepackage{amsmath,amssymb,amsfonts}
\usepackage{algorithmic}
\usepackage{graphicx}
\usepackage{textcomp}
\usepackage{xcolor}
\def\BibTeX{{\rm B\kern-.05em{\sc i\kern-.025em b}\kern-.08em
    T\kern-.1667em\lower.7ex\hbox{E}\kern-.125emX}}
\begin{document}

\title{SpatialVisVR: An Immersive, Multiplexed Medical Image Viewer With Contextual Similar-Patient Search\\
}

\author{
    \IEEEauthorblockN{
        Jai Prakash Veerla\IEEEauthorrefmark{2}\IEEEauthorrefmark{5}, 
        Partha Sai Guttikonda\IEEEauthorrefmark{2}\IEEEauthorrefmark{5}, Amir Hajighasemi\IEEEauthorrefmark{2}\IEEEauthorrefmark{5}, 
        Jillur Rahman Saurav\IEEEauthorrefmark{2}\IEEEauthorrefmark{5},
        Aarti Darji\IEEEauthorrefmark{2}\IEEEauthorrefmark{5},\\
        Cody T Reynolds\IEEEauthorrefmark{2}\IEEEauthorrefmark{5},
        Mohamed Mohamed\IEEEauthorrefmark{2}\IEEEauthorrefmark{5},
        Mohammad S Nasr\IEEEauthorrefmark{2}\IEEEauthorrefmark{5},
        Helen H Shang\IEEEauthorrefmark{2}\IEEEauthorrefmark{4},
        Jacob M Luber\IEEEauthorrefmark{2}\IEEEauthorrefmark{5}}
\vspace{10pt}
    \IEEEauthorblockA{\IEEEauthorrefmark{2}Department of Computer Science, University of Texas at Arlington, USA}
    \IEEEauthorblockA{\IEEEauthorrefmark{5} Multi-Interprofessional Center for Health Informatics, University of Texas at Arlington, USA}
    \IEEEauthorblockA{\IEEEauthorrefmark{4} Ronald Reagan UCLA Medical Center, USA}   
\vspace{10pt}
\IEEEauthorblockA{Email: {jacob.luber@uta.edu}
}
}

\maketitle

\begin{abstract}
% In modern pathology, multiplexed immunofluorescence (mIF) and multiplex immunohistochemistry (mIHC) bring both vast opportunities and challenges. These techniques illuminate complex tumor microenvironment interactions, necessitating intuitive visualization tools to analyze vast biological datasets effectively. With the rise of electronic health records (EHR) and information overload for physicians, integrating advanced technologies becomes essential. Enter SpatialVisVR: a versatile VR platform for comparing medical images, adaptable for data privacy on embedded hardware. Clinicians can capture pathology slides in real-time via mobile, then SpatialVisVR employs a deep learning algorithm to match and display similar mIF images. This interface allows for adding or removing up to 100 multiplexed protein channels, aiding immuno-oncology decisions. Ultimately, SpatialVisVR aims to refine diagnostic processes, promoting a holistic, efficient approach to immuno-oncology research and treatment.

In contemporary pathology, multiplexed immunofluorescence (mIF) and multiplex immunohistochemistry (mIHC) present both significant opportunities and challenges. These methodologies shed light on intricate tumor microenvironment interactions, emphasizing the need for intuitive visualization tools to analyze vast biological datasets effectively. As electronic health records (EHR) proliferate and physicians face increasing information overload, the integration of advanced technologies becomes imperative. SpatialVisVR emerges as a versatile VR platform tailored for comparing medical images, with adaptability for data privacy on embedded hardware. Clinicians can capture pathology slides in real-time via mobile devices, leveraging SpatialVisVR's deep learning algorithm to match and display similar mIF images. This interface supports the manipulation of up to 100 multiplexed protein channels, thereby assisting in immuno-oncology decision-making. Ultimately, SpatialVisVR aims to streamline diagnostic processes, advocating for a comprehensive and efficient approach to immuno-oncology research and treatment.
\end{abstract}

\vspace{5pt}


\maketitle
\begin{figure*}[ht!]
    \centering
        \includegraphics[width=1\linewidth]
        {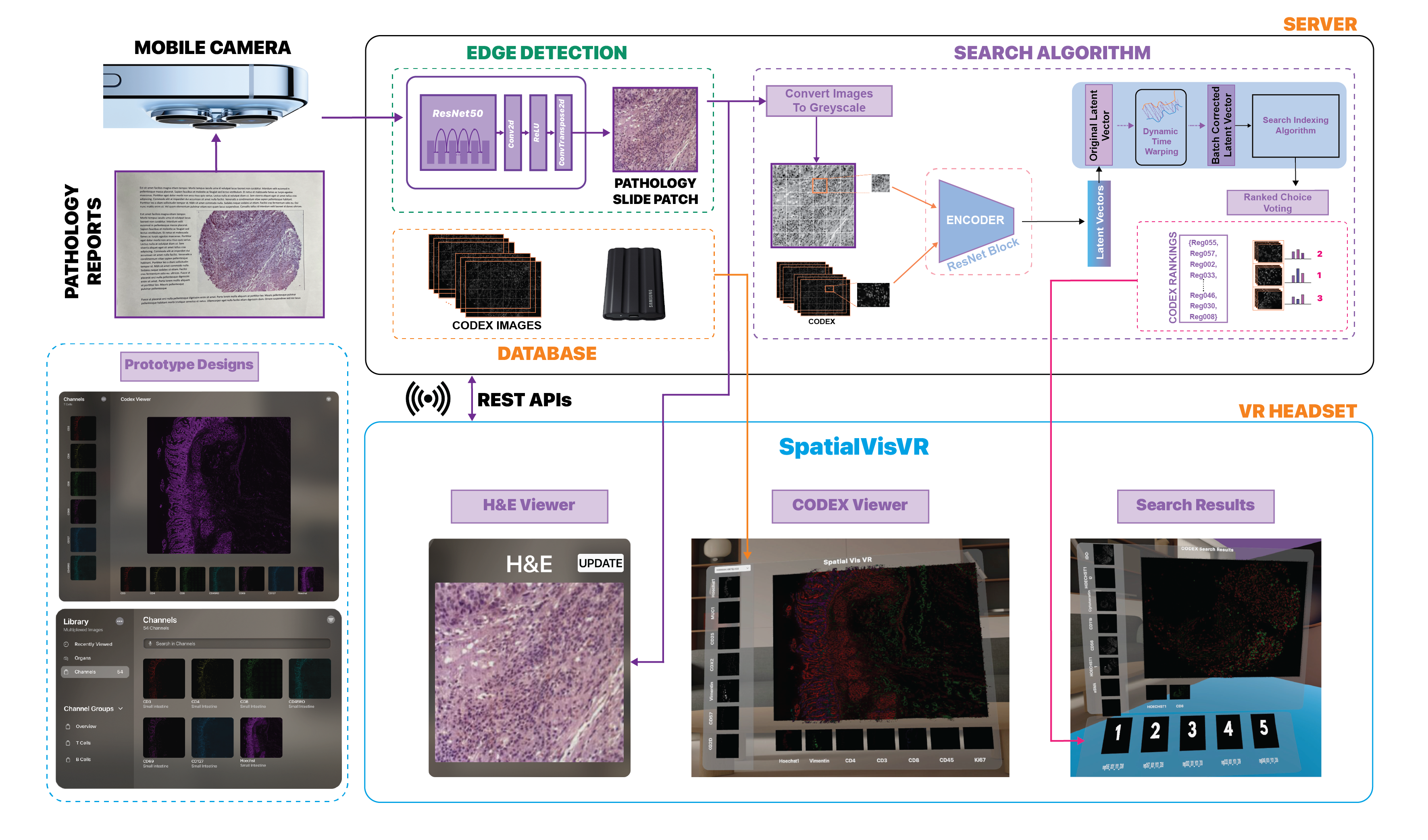}
        \vspace{-20pt}
        \caption{An overview of the SpatialVisVR Pipeline: 1) A medical professional uses the mobile app to take a photo of a pathology image (this could be on a screen, in a textbook, etc.). 2) The Edge detection algorithm captures relevant patches from the slide and streams them to the similar patient search, 3) The similar patient search, which operates by compressing pathology image patches into latent spaces with a variational autoencoder (VAE) and then performing dynamic time warping on these latent spaces, retrieves more clinically useful multiplexed proteomics images that are similar to the query image. Both the machine learning steps will be modular and inference will be able to occur on an ARM Cortex or NVIDIA Jetson nano platform within hospitals, 4) The Unity App in VR streams the original slide the pathologist captured and an interactive viewer where they can select similar multiplexed slides, a comparison viewer to compare between patients, and an interactive viewer to add and subtract multiplexed protein markers to the image.}
        \vspace{-10pt}
        \label{overview}
\end{figure*} 

\section{Introduction}
Multiplexed immunofluorescence (mIF) and multiplex immunohistochemistry (mIHC) are revolutionizing the realm of pathology, especially in the field of immuno-oncology, by providing spatial context for proteomic measurements in whole slide images \cite{harms2023multiplex}. They offer unparalleled insights into the tumor microenvironment and various immune populations \cite{phillips2021highly}. Notably, our SpatialVisVR platform focuses on visualizing CODEX images, which represent state-of-the-art multiplexed tissue imaging technology \cite{goltsev2018deep}. However, these advanced mIF and mIHC images, showcasing approximately 60 markers, present unique visualization challenges \cite{black2021codex}. With approximately 20,000 stainable proteins and each laboratory opting for different markers, standardized visualization becomes challenging. Moreover, most multiplex viewers available today are restricted to desktop or web platforms and typically involve subscription fees \cite{hoffer2020minerva, manz2022viv, stefani2018confocalvr}.

Meanwhile, machine learning approaches play a pivotal role in various areas, ranging from theoretical aspects like fairness defects \cite{monjezi2023information} to practical applications \cite{zadeh2024concrete,esfahani2022application,zare2022experimental}. Additionally, considerable work on different virtual reality (VR) applications is emerging as a transformative tool for data visualization. In the biomedical arena, applications such as visualizing nucleotide sequences and protein structure through BioVR are gaining traction \cite{zhang2019biovr}.

By incorporating Virtual Reality (VR) technology like SpatialVisVR into contemporary pathology, the diagnostic abilities of pathologists can be greatly improved. This is achieved by utilizing the concepts of embodied cognition, which emphasize the need of physically interacting with data for cognitive growth and comprehension. Virtual reality (VR) technologies allow pathologists to easily traverse and edit complicated datasets of multiplexed immunofluorescence (mIF) and multiplex immunohistochemistry (mIHC). This interactive experience surpasses conventional viewing methods, enabling doctors to spatially and dynamically investigate tumor microenvironments using up to 100 multiplexed protein channels. These qualities can enhance their comprehension of complex biological relationships and enhance decision-making in the field of immuno-oncology. Through the utilization of SpatialVisVR, pathologists will be able to not only visually perceive but also actively engage with the data, resulting in a more profound cognitive involvement with the information. This heightened level of engagement is crucial for ensuring precise diagnosis and efficient treatment planning. This method, backed by cognitive research, highlights that cognition is not solely a mental process but also a physical one, where action improves comprehension\cite{10.1145/1142405.1142429}. This makes virtual reality an indispensable tool for pathologists to address the intricacies of contemporary diagnostic procedures.

To facilitate multimodal search SpatialVisVR utilizes Multimodal Pathology Image Retrieval (MPIR) \cite{hajighasemi2023multimodal}, allowing users to efficiently compare medical images. This enables them to promptly get and evaluate related mIF images from a large database for analysis of large biological datasets. This feature offers pathologists a comprehensive perspective on similar patients, assisting in the detection of patterns, abnormalities, and significant biomarkers that are crucial for diagnosing and planning treatment in immuno-oncology. Pathologists can utilize this contextual similarity search to optimize their decision-making processes, extracting knowledge from a wider range of cases and enhancing diagnostic precision. Furthermore, this method simplifies the research process, conserving precious time and resources that would otherwise be utilized in manually seeking out pertinent instances.

In the context of the increasing use of mIF, the dearth of publicly available mIF data becomes evident, especially when compared to the vast archive of traditional H\&E stained images on which many tasks have been performed \cite{shang2023histopathology,robben2023state}. Addressing this disparity, our app, SpatialVisVR, developed in Unity, implements multimodal search using a machine learning approach to visualize similar mIF images (compared to more standard, inputted H\&E slides) that have useful proteomic context for tasks such as selecting cancer immunotherapy treatments. Through a mobile device, one can capture an H\&E slide, subsequently initiating a segmentation to differentiate tissue from the backdrop. The tool then cross-references with its database, pinpointing analogous tissues, thus allowing juxtaposition with mIF-detailed tissues of similar cancer categories. As we advance our tool, we are working towards ensuring its compatibility with lightweight platforms, especially the Jetson Nano. Our aspiration is to offer resource-efficient in-house imaging suitable even for clinics with restricted resources while emphasizing data confidentiality and minimizing transfer risks.

\begin{figure*}[ht!]
    \centering
        \includegraphics[width=0.85\linewidth]
        {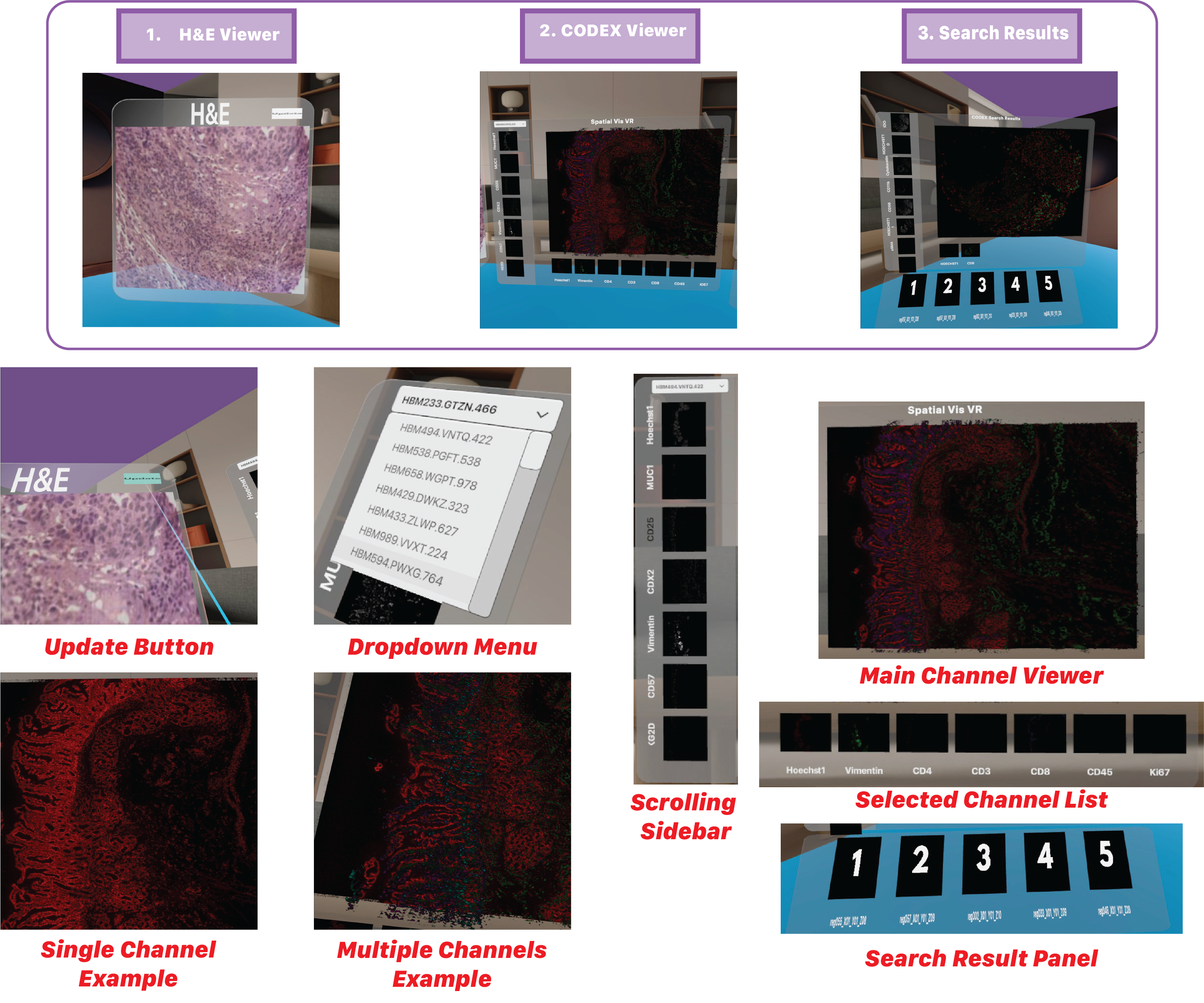}
        % \vspace{-20pt}
        \caption{SpatialVisVR Interface: (Left) H\&E slide visualization; (Center) Multi-channel CODEX ome.tiff viewer with navigation tools; (Right) Top five analogous CODEX images to H\&E slide. Aiding pathologists in diagnostic workflows.}
        \vspace{-10pt}
        \label{frames}
\end{figure*} 

Driven by these trends, our research synergizes VR and mIF to redefine diagnostic paradigms and fortify immuno-oncology research, with a focus on comprehensive quantitative and spatial image examination. In alignment with our mission, we intend to incorporate enhanced quantitative diagnostic tools and delve deeper into cell segmentation or interactions within the cellular environment for more profound insights.

Key contributions of our endeavor include:

\begin{itemize}
\item \textbf{Pioneering VR and mIF Fusion with SpatialVisVR}: To our knowledge, SpatialVisVR is the first VR app geared towards this specialized task, crafting a sophisticated interface for in-depth multiplexed image exploration.
\item \textbf{On-the-Go Mobile Imaging with SpatialVisVR}: Addressing the need for a bridge between traditional H\&E images and mIF data, our software empowers users to effortlessly capture, segment, and cross-reference H\&E slides. Weighing only 53 MB, it serves dual roles as both a multiplexed image viewer and an image search engine.
\item \textbf{Striving Towards an Efficient Pipeline}: In our journey towards fully embracing lightweight platforms, we're taking strides with the Jetson Nano. Our vision is to optimize in-house imaging for all clinics, particularly those with restricted resources, without compromising patient data privacy.
\end{itemize}

Following this introduction, Section 2 discusses related works, Section 3 details our methodologies, Section 4 discusses limitations of the work, section 5 concludes the paper, and Section 6 discusses the future direction of this work.

\begin{figure*}[ht!]
    \centering
        \includegraphics[width=1\linewidth]
        {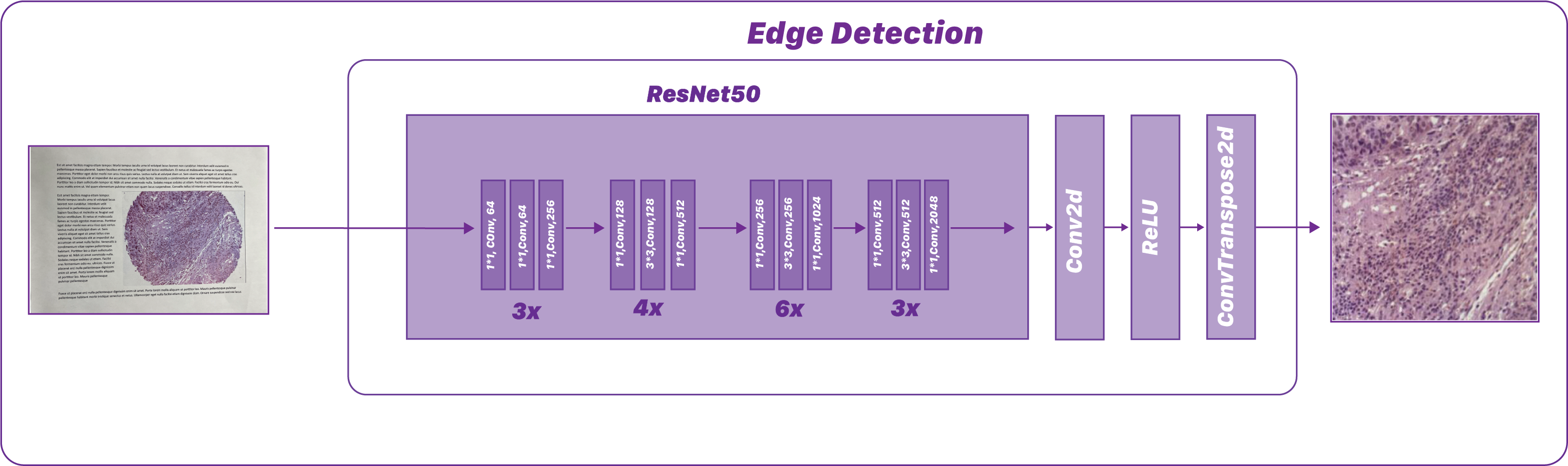}
        \vspace{-20pt}
        \caption{ResNet-50-based detection with a segmentation head for H\&E segmentation from the captured image}
        \vspace{-10pt}
        \label{seg}
\end{figure*} 

\section{Current State of the Field}

%5-7 Paragraphs (Do not use chatGPT for this as it will make things up, I am fine with you guys using chatGPT to clean up other parts of the manuscript as long as it is factually correct; cite at least 20-25 prior works to justify why ours is novel. This needs to include H\&E, CODEX, VR, AR papers)

% I have structured the lit review for a connected story, old one is available at LR_Old.txt - Saurav

In the burgeoning field of medical research, emerging technologies like Virtual Reality (VR) and Augmented Reality (AR) are gaining prominence due to their potential in revolutionizing clinical practice and medical education. Our SpatialVisVR sits at the intersection of multiplexed imaging, machine learning in pathology, and the immersive capabilities of VR and AR, highlighting the advancements and challenges in these interconnected domains.

Histopathology has traditionally relied on H\&E stained imaging for cellular and tissue structure visualization \cite{bejnordi2015stain}. However, modern techniques like CODEX have driven a paradigm shift, enabling visualization of over 60 protein markers within a single tissue site, thus enhancing cellular-level understanding and facilitating sophisticated data amalgamation in complex fields such as oncology \cite{black2021codex, boehm2022harnessing, cheerla2019deep}.

The research conducted by Veerla et al. \cite{veerla2023analyzing} highlighted the intricate nature of analysing extensive biological datasets, pushing us to recognize the significance of efficient visualization tools. Their research emphasized the necessity of strong visualization methods to assist in the examination of complex transcriptome and proteome data, which influenced our strategy in creating tools such as SpatialVisVR.

The complexity of navigating through advanced imaging data, particularly Whole-Slide Imaging (WSIs), which can be up to 50k by 50k pixels in size, presents a unique challenge \cite{yagi2012ultra}. Emerging solutions such as Mistic, Viv, and Minerva address this challenge by providing intuitive navigation and efficient data access for multiplexed imaging, utilizing client-side GPU rendering and cloud-supported guided analyses \cite{manz2022viv, hoffer2020minerva}.

VR and AR play a pivotal role in modern medical imaging, with applications ranging from virtual surgeries to diagnostic advancements \cite{javaid2020virtual, farahani2016exploring}. Innovations like ConfocalVR and ExMicroVR showcase VR's potential in rendering cellular structures, while AR continues to evolve in therapeutic and educational domains despite challenges in achieving optimal visual fidelity \cite{stefani2018confocalvr, cheng2023micromagnify, ulrich2022ardiff}.

The integration of traditional histopathology, advanced multiplexed imaging, and immersive tools like VR and AR heralds a new era in pathology imaging. Platforms such as Minerva and Avivator lead the way in rendering multiplexed visuals with unparalleled detail, envisioning a future where VR and AR significantly enhance the user experience in exploring both traditional and multiplexed images \cite{hoffer2020minerva, keller2021vitessce}.

\section{Methods}

\begin{figure*}[ht!]
    \centering
        \includegraphics[width=1\linewidth]
        {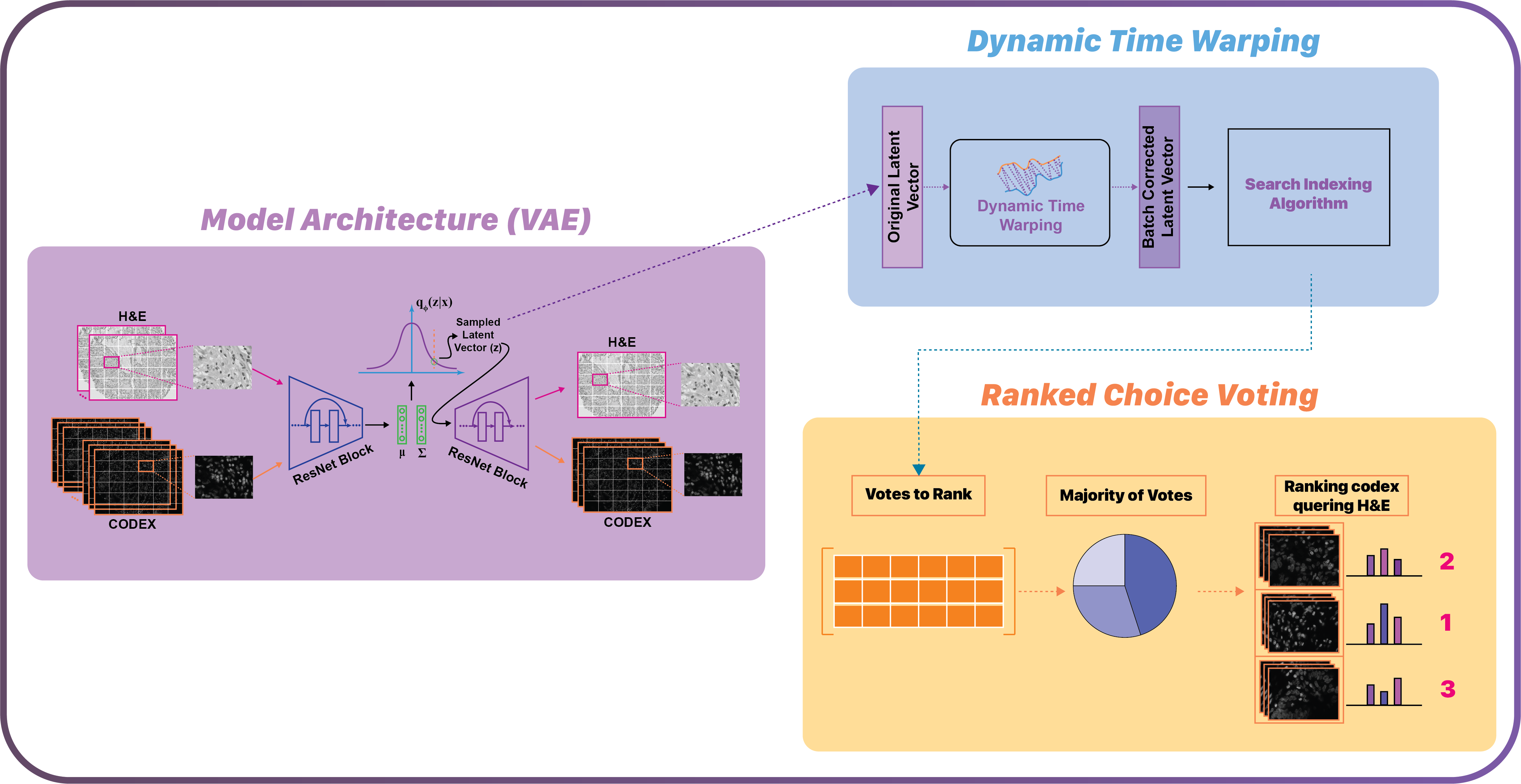}
        \vspace{-20pt}
        \caption{Overview of the search engine: 1) Multimodal VAE Architecture compresses H\&E and mIF images. 2) Dynamic Time Warping to integrate latent space of mIF and H\&E patches. 3) Using cosine similiarity and centroid based indexing to compare latent spaces accross image modalities 4) Ranked Choice voting to aggregate patch level similiarty to slide level similarity}
        \vspace{-10pt}
        \label{vae}
\end{figure*} 

This section outlines how we utilized methodologies, including object detection, hardware setups, and Multimodal Pathology Image Retrieval (MPIR) integration, for enhanced pathology data analysis and visualization, linking computational techniques with tangible user interactions.

\subsection{SpatialVisVR App}
\subsubsection{VR Application:}
We utilized VR to visualize CODEX images as intricate 3D channel stacks. On the server-side, we implemented Flask to cache user-selected CODEX images in RAM for faster data retrieval. We created APIs to return channel names and the nth channel of specified CODEX images, enabling dynamic rendering in VR. We facilitated interactive coloring selection via APIs, where we processed user-selected channels server-side based on color and threshold parameters before sending them to VR for 3D arrangement and visualization, enhancing them with Unity shaders for better clarity and color vibrancy.\\

% \vspace{10pt}
\subsubsection{GUI Overview:}
SpatialVisVR, designed for pathologists, focuses on providing a seamless user experience that utilizes spatially multiplexed data to improve the diagnostic workflow, as showcased in Fig. \ref{vae} The interface comprises three tailored windows that serve distinct purposes in the diagnostic workflow.

The left window visualizes H\&E slides through the edge detection of real-time captures, featuring an 'Update' button to initiate searches in the third window and a primary viewing area for real-time image processing.

The central window, serving as the primary viewer, facilitates the exploration of CODEX ome.tiff images, allowing users to view up to seven channels simultaneously for understanding cellular interactions. Key UI elements include a sliding carousel for channel selection, a dropdown for accessing stored CODEX images, and a navigation bar for channel management, with each channel color-coded for differentiation.

The right window displays search results for similar CODEX images to an H\&E slide, showing the five most similar CODEX images and aiding in further exploration. 

Overall, the multi-window setup in SpatialVisVR enhances pathologists' workflow and diagnostic decision-making by providing diverse data examination, critical insights, and a more efficient diagnostic process. An overview of the interface depicted in Fig. \ref{frames}.\\

% \vspace{10pt}
\subsubsection{API Implementation}
For communication between the server-side application and VR, we deployed a server-client model using RESTful APIs \cite{8385157}. We set up a Flask application on the server end and used Unity to process API call data, enhancing data visualization with shaders. To improve user experience and address VR headset limitations, we shifted resource-intensive tasks to the server. The server, equipped with a 4TB SSD, hosted CODEX datasets, and limited the data transfer rate to 200 MB for better VR responsiveness.

\subsection{ML Backbone}

\subsubsection{Object Detection Implementation}
The object detection system utilized a ResNet-50 backbone implemented using PyTorch, wherein the classification layer was replaced with a segmentation head composed of a Conv2D layer, ReLU layer, and ConvTranspose layer. This configuration processed the ResNet-50 output through a 3x3 convolution operation and scaled the image to a 2x640x640 output using ConvTranspose2D, thereby distinguishing between pathology slide and non-pathology slide pixels. The architecture, depicted in Fig. \ref{seg}, illustrates this setup.

During the training phase, image preprocessing included Resize, RandomHorizontalFlip, RandomInvert, RandomRotation, GaussianBlur, and ColorJitter transforms to augment model robustness. For prediction, preprocessing steps involved resizing the images to 640x640, applying padding, and subsequent resizing to rectify bounding area inconsistencies.

Performance was influenced by various aspects of photo setup such as lighting conditions, angle of capture, and focus. Consistent lighting conditions resulted in superior outcomes, while shadows or inconsistent lighting led to suboptimal results. The system performed optimally at 90-degree angles, with acceptable outcomes at shallow angles. However, steep angles or out-of-focus photographs resulted in inconsistent detection and bounding, impacting the overall output quality.\\

\subsubsection{MPIR Implementation}
We utilized Multimodal Pathology Image Retrieval (MPIR) to retrieve similar codex images to a given H\&E image slide \cite{hajighasemi2023multimodal}. The process involved encoding images into lower-dimension latent vectors using a Variational Autoencoder (VAE). This VAE, comprising an encoder and decoder, mapped input data to a probabilistic latent space and generated data samples from these latent variables, aiding in dimensionality reduction and data reconstruction \cite{nasr2023clinically, vora2023real}.

Initially, we trained the VAE on H\&E images, followed by separate training on seven specific channels of codex images selected for their relevance in indicating tumor and immune cell interactions. Preprocessing included selecting useful patches from high-resolution images for training purposes. During testing, we obtained latent vectors from codex and H\&E image patches, using the dynamic time warping algorithm \cite{1163055} for batch effect correction. Subsequently, we employed cosine similarity and a search algorithm to identify the top 5 most similar patches of codex slides for each H\&E image patch. Results across different channels were consolidated using rank choice voting algorithms to retrieve the most similar codex slides for the H\&E input. This algorithm was also applied to camera-captured H\&E images post edge detection processing.

To test the MPIR model, we used two H\&E image inputs: the actual image and the result from the edge detection algorithm. For the actual image, we obtained the top five similar codex slides, and a different set was identified for the edge-detected image, with each slide assigned a voting number representing its similarity rank.

% \subsubsection{MPIR Input Comparison}

\begin{figure*}[h!]
    \centering
        \includegraphics[width=1\linewidth]
        {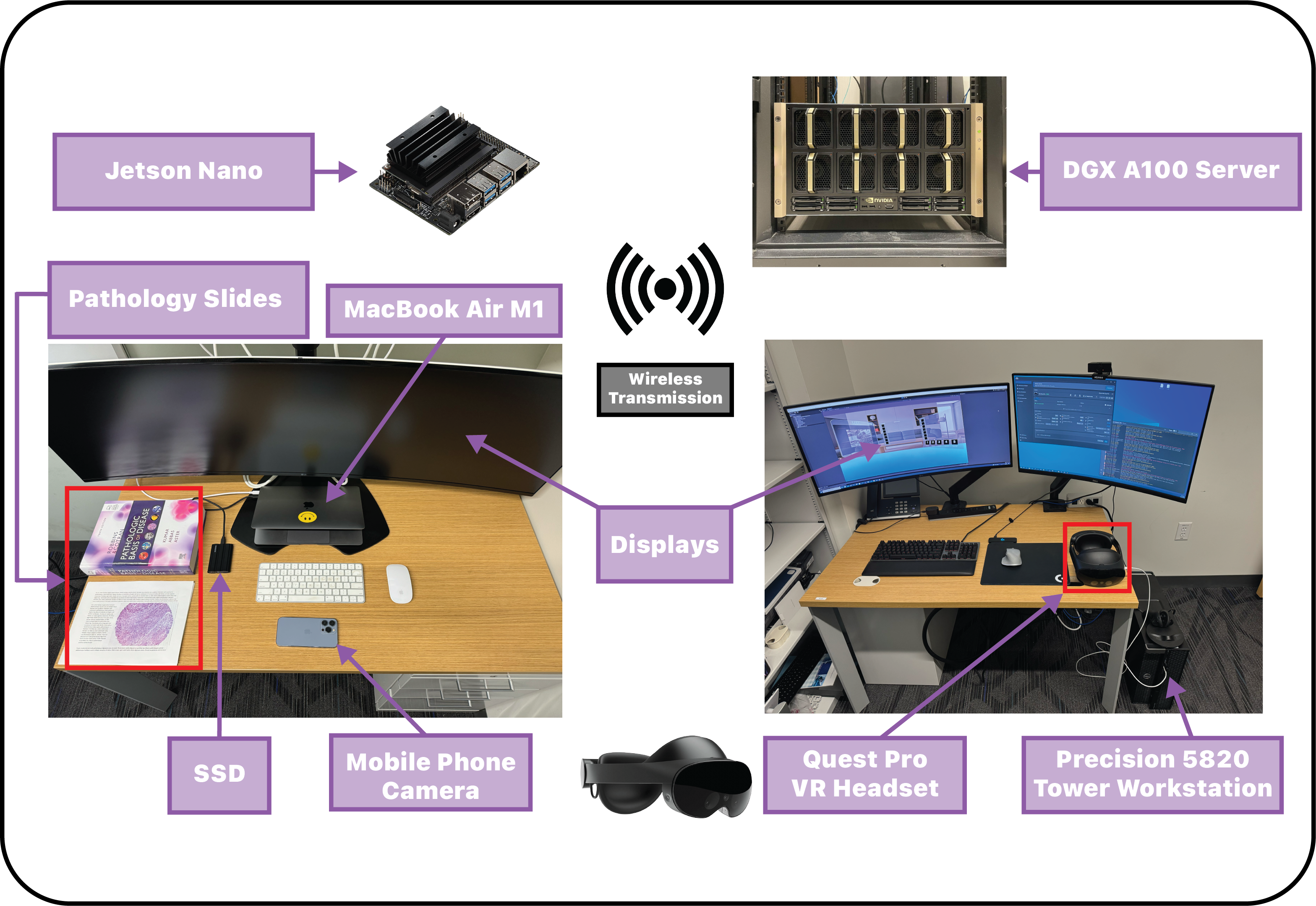}
        % \vspace{-20pt}
        \caption{Hardware Overview: Smartphones for H\&E capture, Jetson Nano and MacBook for detection, DGX Server, Precision tower for MPIR training, and Meta Quest Pro for visualization.}
        \vspace{-10pt}
        \label{hardware}
\end{figure*}

\subsection{Hardware Setup}
We utilized various hardware components for this project. iOS and Android smartphones were employed to capture H\&E slide images in real-time, functioning as IP Webcams over a Local Area Network (LAN). The OpenCV Python library facilitated video stream interfacing due to its GStreamer integration and object detection capabilities.

For testing the object detection model, we utilized a Jetson Nano 2GB Developer Kit to ensure real-time performance. Deployment tasks were carried out using an Apple 13-inch MacBook Air with an M1 chip and 16GB RAM. Post-object detection, a DGX Server equipped with eight NVIDIA A100 GPUs was used for testing and training the MPIR algorithm. The execution phase was then shifted to the MacBook Air for practical purposes.

During each inference run with our model, the top 5 candidates identified by the MPIR's ranked choice voting algorithm were transmitted to a Meta Quest Pro headset via a Unity Development Environment on a Dell Precision 5820 Tower Workstation for final visualization. After viewing the visualization, the physicians on our team offered valuable feedback on the user interface design. Looking ahead, our future work aims to fully deploy the pipeline exclusively to Jetson Nano, ARM Cortex, and any compatible VR headset. The hardware setup is depicted in Fig. \ref{hardware}, providing an overview of the different components utilized in this research.

\subsection{Data Used}
We trained the VAE model using H\&E and mIF images from CODEX technology, which included colorectal cancer samples from 35 patients with over 50 protein markers. These images were captured using a Keyence BZ-X710 microscope \cite{schurch2020coordinated}. Additionally, we visualized 109 stitched CODEX images from NIH HuBMAP, representing various organs, in VR \cite{hubmap2019human}. For object detection, we utilized a dataset comprising around 500 H\&E slides from three universities and 195 H\&E slides from The Cancer Genome Atlas for breast cancer. These slides were digitized using Aperio or Ventana scanners \cite{cruz2018high}. 

\section{Discussion} 
SpatialVisVR represents a significant advancement in the visualization of multiplexed immunohistochemistry slides using virtual reality. However, the transfer of large multiplexed imaging datasets through APIs imposes a considerable burden on the machines processing these images, potentially affecting performance and usability. Additionally, while VR offers an immersive experience, prolonged use can lead to discomfort, although this is mitigated by incorporating user-centered design tailored to pathologists' needs. Moreover, deploying such technology in resource-constrained settings poses challenges due to high computational demands. Furthermore, the Multimodal Pathology Image Retrieval (MPIR) engine \cite{hajighasemi2023multimodal}, a core feature of SpatialVisVR, requires higher quality H\&E images as input since lower resolution images might not yield good search results. Being the first of its kind to search and query multiplexed proteomics imaging such as CODEX from H\&E stained images, the algorithm's similarity search results may not always align with clinical expectations due to the limited data available for training. Future iterations of the platform will need to focus on optimizing data handling, improving search accuracy, and enhancing user comfort to extend its applicability and effectiveness in diverse clinical environments.

\section{Conclusion} 

Traditional pathology primarily relies on immunohistochemistry staining for diagnostic or prognostic evaluations. Multiplex immunohistochemistry staining has enriched insights on spatial cellular interactions, especially in oncology, elucidating tumor microenvironments (TME) \cite{black2021codex}\cite{allam2022spatially}\cite{schurch2020coordinated}. However, the 2D nature of multiplexed images constrains spatial exploration to the x and y axes. Recent attempts towards 3D reconstructions of multiplexed images will further spatial understanding of disease processes, yet 2D interfaces for visualization remain a challenge \cite{lin2023multiplexed}.

Our VR platform addresses this by enabling interactive visualization of multiplexed immunohistochemistry stains that are contextual to the environment of existing H\&E slides, facilitating intuitive exploration for researchers and clinicians. This capability enhances the analysis of spatial features in medical imaging and provides a more interactive environment for navigating whole slide images in the context of clinically richer multiplexed immunofluorescent ones, aiding in better diagnostic and prognostic evaluations. 

The incorporation of Virtual Reality (VR) technology with digital pathology will enable pathologists to not only visually perceive the data but also actively engages with the data, leading to a better cognitive involvement with the information. This added ability of involving cognition while analyzing the data with an extension to actions is something that cannot be achieved by the tradition way of interacting via 2D displays.  Furthermore, our platform facilitates slide comparison and retrieval, directing pathologists towards potential diagnoses or treatment responses based on observations from similar patients by other clinicians.

\section{Future Directions}

We aim to enhance the resolution of multiplexed slides in VR to display distinct cellular interactions. This advancement could enable manual manipulation of cellular interactions and prediction of downstream effects within VR. 

Integration of features for swift extraction of diagnostic or prognostic information is also envisioned. For instance, facilitating quick calculations of HER2 expression in 3D to ascertain candidacy for HER2-directed therapies. We plan to broaden slide comparison and retrieval capabilities by expanding our spatial atlas of multiplex slides with contributions from end-users. 

Transitioning towards exploring entire tumors in 3D, beyond individual multiplexed slides, will allow easy visualization of protein expression heterogeneity like HER2 or PDL1 in VR. This could uncover missed treatment opportunities from single biopsy slides, potentially impacting treatment avenues for millions globally.

\section{Demo Video, Code and Data Availability}
A demonstrative video by the authors is available in the GitHub Repository summarizing the method's potential benefits for pathologists. A version of our code sufficient to recreate our entire pipeline is available on GitHub at: https://github.com/jaiprakash1824/SpatialVisVR. All the Multiplexed imaging datasets used are available from the NIH HuBMAP consortium at: https://hubmapconsortium.org.

\section*{Acknowledgment}

This work was supported by a University of Texas System Rising STARs Award (J.M.L) and the CPRIT First Time Faculty Award (J.M.L)

\bibliographystyle{IEEEbib}
\bibliography{refs}

\end{document}